\documentclass[sigconf]{acmart}

\renewcommand\footnotetextcopyrightpermission[1]{} % removes footnote with conference information in first column

\setcopyright{none}

\acmConference[SE4FP 2026]{2nd Workshop on Software Engineering for Functional Programming}{September 8, 2026}{São Paulo, SP, Brazil}

\AtBeginDocument{
    
}

\usepackage{orcidlink}
\usepackage{listings}
\usepackage{xcolor}
\usepackage{longtable}
\usepackage{array}
\usepackage[most]{tcolorbox}

\definecolor{codebg}{RGB}{245,245,245}
\definecolor{querykw}{RGB}{0,0,150}

\begin{document}

\newcommand{\orcidauthorA}{0009-0005-3988-5162}

%% allowing the author to define a "short title" to be used on the page 
%% headers.
\title{Code Smells in Clojure: Initial Findings from a Grey Literature Review}

\author{Walber Araújo}
\affiliation{
  \institution{Federal University of Campina Grande (UFCG)}
  \city{Campina Grande}
  \country{Brazil}
}
\email{walber.araujo@splab.ufcg.edu.br}

\author{José Truta}
\affiliation{
  \institution{Federal University of Campina Grande (UFCG)}
  \city{Campina Grande}
  \country{Brazil}
}
\email{jose.truta@splab.ufcg.edu.br}

\author{Lucas Vegi}
\affiliation{
  \institution{Federal University of Viçosa (UFV)}
  \city{Viçosa}
  \country{Brazil}
}
\email{lucas.vegi@ufv.br}

\author{Marco Tulio Valente}
\affiliation{
  \institution{Federal University of Minas Gerais (UFMG)}
  \city{Belo Horizonte}
  \country{Brazil}
}
\email{mtov@dcc.ufmg.br}

\author{João Brunet}
\affiliation{
  \institution{Federal University of Campina Grande (UFCG)}
  \city{Campina Grande}
  \country{Brazil}
}
\email{joao.arthur@computacao.ufcg.edu.br}

%% By default, the full list of authors will be used on the page
%% headers. This list is often too long and will overlap
%% other information printed in the page headers. 
%% This command allows the author to define a more concise list
\renewcommand{\shortauthors}{Araújo et al.}
\renewcommand{\shorttitle}{Code Smells in Clojure: Initial Findings from a Grey Literature Review}

\begin{abstract}
Code smells are widely used indicators of poor code quality, revealing structural problems and areas where improvement can be made. Although extensively studied in object-oriented languages, functional programming languages remain comparatively underexplored in literature. This paper presents early results from a grey literature investigation of code smells in Clojure, a modern functional programming language that runs on the Java Virtual Machine (JVM) and is widely adopted in open source and industrial systems. Inspired by prior work on Elixir, we manually inspected developer discussions retrieved through Google search, extracted quality concerns discussed by developers, and had 44 practitioners evaluate the relevance of non-traditional smell candidates. As preliminary results, we cataloged 26 code smells, including 12 Clojure-specific, 9 functional-style, and 5 traditional Fowler smells. We also analyzed tool support and observed a significant gap, as existing Clojure linters cover only 2 of these 26 smells. These early findings provide an initial characterization of how Clojure developers discuss code smells, a preliminary set of smell-like problems specific to this ecosystem, and an early assessment of tool support for their detection.
\end{abstract}


%% This command processes the author, affiliation, and title
%% "frenchspacing" avoids an additional space after a period at the end of a sentence.
\frenchspacing
\maketitle

\section{INTRODUCTION}
\label{sec:intro}

As software systems evolve, ensuring quality aspects that positively influence maintainability becomes increasingly important, especially because development costs is typically associated with maintenance activities~\cite{brooks1982,1265817}. Poor design or implementation decisions may introduce recurring symptoms of degraded quality, commonly referred to as code smells, which may reveal deeper problems and indicate opportunities for refactoring~\cite{fowler1999}. Although they do not necessarily affect program execution, such problems may hinder maintenance and software evolution over time. The foundational catalog by Fowler and Beck has motivated extensive research on code smells across diverse domains, from mobile applications~\cite{Carvalho2019Empirical} to machine learning systems~\cite{Zhang2022Code}. These studies demonstrate that smells establish a shared vocabulary for quality issues, directly influencing the development of static analysis tools like SonarQube~\cite{sonarqube} and PMD~\cite{pmd} to automatically detect such problems.

While code smells have been widely studied, functional programming languages remain comparatively underexplored in this literature. Prior work on Elixir provided early evidence that developers discuss both traditional smells and language-specific quality concerns~\cite{Vegi_2022}. This line of research was later extended into broader investigations of Elixir-specific code smells and refactorings, validated with developers from the community~\cite{francisco2023understanding,ctd}. However, these findings are still limited to a single ecosystem. Moreover, the authors explicitly point to the investigation of other modern functional languages, such as Clojure~\cite{clojure_official}, as a direction for future work. Motivated by this gap, we replicate the original study in the Clojure ecosystem to investigate whether similar discussions about code smells also emerge in this context.

In this paper, we present early results from a grey literature investigation of code smells in Clojure, a modern functional programming language that runs on the Java Virtual Machine (JVM) and is adopted in industrial systems by companies such as Apple and Nubank\footnote{\url{https://clojure.org/community/companies}}. Despite its growing popularity in industry and its strong ecosystem interoperability with Java, research on software quality concerns in Clojure remains comparatively limited. Inspired by previous work on Elixir, we analyze discussions retrieved through Google search to investigate whether developers mention traditional smells, discuss other smell-like problems, and whether such issues are supported by a well-known static analysis tool. Since this study focuses on a single source of grey literature, our goal is not to provide a validated catalog, but to report preliminary evidence on smells and related quality concerns in developer discussions.

To conduct our replication, we adapted the original research questions to focus on the Clojure ecosystem:

\begin{itemize}
    \item RQ1. Do \textit{Clojure} developers discuss traditional code smells?
    \item RQ2. Do \textit{Clojure} developers discuss other smells?
    \item RQ3. Does a well-known static code analysis tool for \textit{Clojure} detect code smells?
\end{itemize}

To answer these questions, we manually inspected the retrieved grey literature documents, selected those relevant to our research questions, and extracted the smells discussed by developers. We then distinguished traditional smells from non-traditional candidates treated as recurring quality concerns in this ecosystem. After an initial refinement step, practitioners evaluated the relevance of these candidates based on their names and descriptions. Finally, we examined whether a well-known static analysis tool detects them.

The main contributions of this paper are: (i) an initial characterization of how Clojure developers discuss traditional code smells in grey literature; (ii) a preliminary set of other smell-like problems discussed in the Clojure ecosystem; and (iii) an early assessment of the extent to which a Clojure static analysis tool supports the detection of these problems.

The remainder of this paper is organized as follows. Section~\ref{sec:background} presents background and related work. Section~\ref{sec:methodology} describes the methodology we employed to conduct our work. Then, in Section~\ref{sec:results}, we report the results for each research question. Section~\ref{sec:threats} discusses threats to validity. Section~\ref{sec:future-work} concludes the paper and outlines future work. Finally, Section~\ref{sec:replication-package} presents the replication package.

\section{BACKGROUND AND RELATED WORK}
\label{sec:background}

\subsection{Background}

\textbf{Clojure.} Clojure is a modern Lisp dialect~\cite{lisp} designed as a practical functional programming language~\cite{hickey2008clojure}. The language emphasizes immutable data structures, first-class functions, macros, and data-oriented programming. Clojure programs are commonly organized into namespaces, which group related definitions and manage dependencies between modules. Functions are defined as immutable values and are typically composed through higher-order operations and sequence-processing abstractions. The language also provides a homoiconic syntax, in which programs are represented as native data structures rather than character streams~\cite{clojure_reader}, supporting code manipulation and metaprogramming through macros. In recent years, Clojure has also gained industrial adoption in companies such as Amazon, Nubank, Netflix, Apple, Spotify, Facebook, and Walmart Labs\footnote{\url{https://clojure.org/community/companies}}.

Many of these design choices were influenced by previous research on software complexity, persistent data structures, and concurrency\footnote{\url{https://clojure.org/about/documentary}}. In particular, Clojure adopts persistent collections inspired by Hash Array Mapped Tries (HAMTs)~\cite{bagwell2001ideal}, and its Software Transactional Memory (STM) model was influenced by prior work on composable memory transactions~\cite{harris2005composable}. Additionally, the language design was influenced by discussions on reducing accidental complexity associated with mutable state in software systems~\cite{mosley2006out}. These characteristics shape how Clojure programs are structured and maintained.

\textbf{Code Smells.} Code smells are recurring structures in source code that suggest weaknesses in a system's internal design and may indicate the need for refactoring. The term was popularized by Fowler and Beck~\cite{fowler1999}, who used it to describe code fragments that are not necessarily incorrect, but may make a system more difficult to comprehend, modify, or extend. In this sense, they provide a practical vocabulary for discussing internal quality problems and for identifying parts of the code that deserve further inspection.

Fowler and Beck's catalog was originally shaped by object-oriented software and includes smells such as \textit{Duplicated Code}, \textit{Long Method}, and \textit{Long Parameter List}. However, the way smells appear in practice may depend on the language, paradigm, and development conventions adopted in a project. While some smells can be observed across different ecosystems, others are closely tied to specific constructs or idioms. In this paper, we use the term \textit{traditional code smells} to refer to smells from Fowler and Beck's original catalog.

%\textcolor{red}{precisamos de alguma coisa que apoie a frase "main static analysis tools."}
\textbf{Static Analysis in Clojure.} The ecosystem provides several tools for improving code quality, including linters, formatters, and editor integrations. In this work, we consider widely used static analysis tools in the Clojure ecosystem, namely \textit{clj-kondo}~\cite{clj_kondo}, \textit{Eastwood}~\cite{eastwood}, \textit{Joker}~\cite{joker}, and \textit{Kibit}~\cite{kibit}. These tools are commonly used to report potential errors, warnings, and style-related issues without executing the program, providing continuous feedback during development.

Among these tools, \textit{clj-kondo} is widely adopted in practice, particularly due to its integration with modern development workflows. It is often used in combination with \textit{clojure-lsp}~\cite{clojure_lsp}, which exposes its diagnostics through the Language Server Protocol, enabling editor support such as Visual Studio Code via Calva\footnote{\url{https://calva.io/}}. This widespread integration makes \textit{clj-kondo} a representative example of how static analysis tools are used in everyday Clojure development.

\subsection{Related Work}

Although code smells have been widely studied, few works focus on functional programming languages. Some studies have investigated quality problems or refactoring support in functional settings, especially in languages such as Erlang and Haskell. For example, Cowie~\cite{haskellSmellsCowie2005} proposed smell detection for Haskell programs, and Li and Thompson~\cite{ErlangModularitySmells} identified modularity smells in Erlang systems. Other studies examined duplicated code in Erlang~\citep{CodeCloneErlang04,CodeCloneErlang01,CodeCloneErlang02,CodeCloneErlang03} and Haskell~\citep{CodeCloneHaskell01}. However, to the best of our knowledge, no previous work has investigated code smells in the Clojure ecosystem.

The closest work to ours is the study by Vegi and Valente on code smells and refactoring strategies in Elixir. Their initial study used a grey literature review to investigate whether Elixir developers discuss traditional smells, whether they mention Elixir-specific smells, and whether such smells are detected by Credo, a static analysis tool for Elixir~\cite{Vegi_2022}. This line of work was later extended into a broader catalog of Elixir smells, validated with experienced developers~\cite{francisco2023understanding}. More recently, Vegi and Valente conducted a thesis that expanded this line of research by cataloging both code smells and refactoring strategies for Elixir, as well as establishing relationships between them to support the disciplined removal of smells through refactoring actions~\cite{ctd}.

Our work follows the motivation of this line of research but focuses on Clojure, providing early evidence based on Google-retrieved grey literature, initial practitioner feedback, and an analysis of tool support.

\section{METHODOLOGY}
\label{sec:methodology}

This study adapts the grey literature review conducted by Vegi and Valente for Elixir~\cite{Vegi_2022} to Clojure. We preserve the main steps of their early study, including Google search, document selection, data extraction, and the assessment of static analysis support. In addition, we extend the process with an initial interaction with the Clojure community and a practitioner survey to collect feedback on the relevance of the smell candidates. These steps were designed to answer the research questions presented in Section~\ref{sec:intro}.

Figure~\ref{fig:methodology} summarizes our methodology. To answer \verb|RQ1| and \verb|RQ2|, we took the following steps:

\begin{figure}[htbp]
    \centering
    \includegraphics[width=\linewidth]{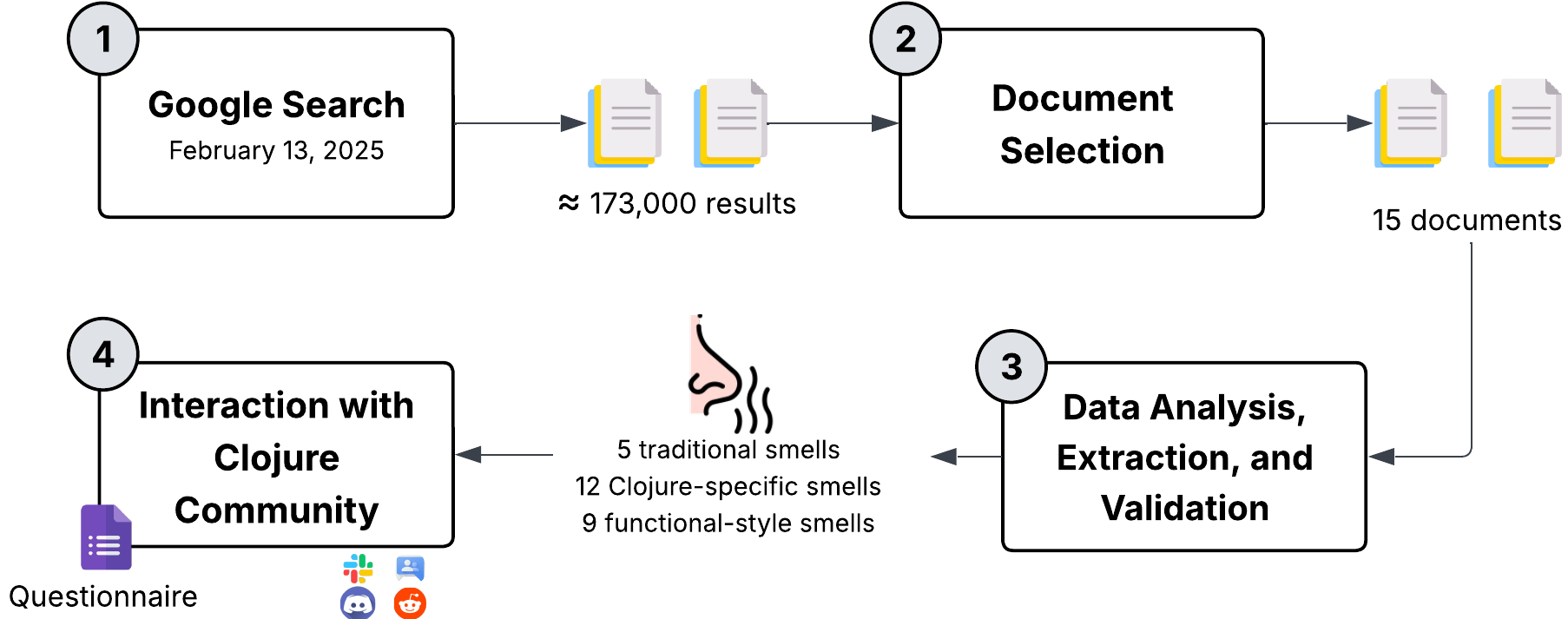}
    \caption{Overview of the research methodology.}
    \label{fig:methodology}
\end{figure}

\paragraph{1) Google Search:} Following the methodology of the original Elixir study~\cite{Vegi_2022} and recommendations for grey literature reviews~\cite{garousi2019}, we used Google Search to retrieve grey literature documents discussing code smells in Clojure. The search was manually conducted on February 13, 2025, using the query shown in Listing~\ref{lst:search-query}, which preserves the original query structure while replacing only the programming language name. The 60 analyzed documents are available in the \textit{Google Top-60.xlsx} spreadsheet included in the replication package~\cite{replication-package}.

\begin{lstlisting}[caption={Google search query used in the study},label={lst:search-query}]
("Clojure") AND
("code smell" OR "code smells" OR "bad smell" OR 
 "bad smells" OR "anti-pattern" OR "anti-patterns" OR
 "antipattern" OR "antipatterns" OR "bad-practice" OR 
 "bad-practices" OR "bad practice" OR "bad practices")
\end{lstlisting}

\paragraph{2) Document Selection:} The search returned approximately 173,000 results. Since inspecting all retrieved pages would be impractical, grey literature reviews commonly restrict screening to the most relevant search results~\cite{garousi2019}. Therefore, using the same criteria of the original study, we manually analyzed the top 60 results returned by Google. The first two authors independently screened these documents and selected those relevant to our research questions, i.e., documents containing discussions about problematic, non-idiomatic, or quality-related practices in Clojure.

After this step, 15 documents were selected for further analysis. The remaining results were excluded because they did not discuss code quality concerns in Clojure or only aggregated external links. No relevant documents were identified among the last ten inspected results, suggesting lower relevance in lower-ranked pages.

For example, the Reddit discussion ``Functional programming anti-patterns?'' (G1), published in the \textit{r/Clojure} community, was selected because practitioners discussed candidate smells aligned with our study, including \textit{Long Method}, \textit{Long Parameter List}, \textit{Deeply-nested Call Stacks}, \textit{Overabstracted Composition}, \textit{Overuse of Higher-order Functions}, \textit{Explicit Recursion}, and \textit{Namespaced Keys Neglect}.

%\textcolor{red}{trazer um exemplo curto de documento que foi selecionado e o porquê.}
\paragraph{3) Data Extraction and Validation:} The first two authors independently examined the selected documents to identify excerpts relevant to RQ1 and RQ2. Once this stage was completed, both assessments were compared and reviewed jointly in order to reconcile disagreements and improve consistency in the extracted evidence. Only two conflicts emerged during this procedure. After discussion, the smell (\textit{Unnecessary Macros}) was retained, whereas the document identified as \verb|G8| in the replication package was excluded for not being aligned with the research questions.

\paragraph{4) Interaction with Clojure Community:} Between July and September 2025, we publicized the study across major Clojure community channels, including Slack, Discord, Google Groups, and Reddit, listed on the official Clojure community resources page\footnote{\url{https://clojure.org/community/resources}}. Initial contacts were also established through issues in the GitHub repository where we were cataloging the smells, from which we received early feedback indicating that traditional code smells were generally perceived as less relevant in the Clojure ecosystem. Community members also suggested conducting a survey to assess the perceived relevance of the identified smells. Based on this feedback, we designed and conducted a practitioner survey containing all candidate smells identified in this study, except for traditional smells, which were removed after the initial community feedback. For each smell, participants could optionally indicate whether they considered it relevant (\textit{Yes}, \textit{No}, or \textit{I don't know}) and provide additional comments. The survey received 44 responses and 214 optional comments.

In order to answer RQ3, we analyzed the official documentation of the most widely adopted static analysis tools in the Clojure ecosystem, namely \textit{clj-kondo}, \textit{Eastwood}, \textit{Joker}, and \textit{Kibit}, to determine which traditional code smells and additional smells identified in the Clojure ecosystem are supported by their rules and diagnostics. The identified checks were subsequently mapped to the smells discussed by developers in our dataset.

\section{RESULTS}
\label{sec:results}

This section presents the results obtained in this study. The findings are organized according to the research questions. First, Section~\ref{sec:rq1-result} investigates whether Clojure developers discuss traditional code smells. Next, Section~\ref{sec:rq2-result} analyzes whether developers discuss other types of smells specific to the language and its ecosystem. Finally, Section~\ref{sec:rq3-result} examines whether a well-known static code analysis tool for Clojure detects code smells reported by developers.

\subsection{RQ1. Do Clojure developers discuss traditional code smells?}
\label{sec:rq1-result}

Although traditional code smells are discussed in the analyzed documents, the terminology introduced by Fowler is not explicitly used in some cases. Instead, the issues are described in terms of their characteristics, as illustrated in the examples below.

The \textit{Long Parameter List} smell was the only one observed in more than one document (\verb|G1| and \verb|G2|), associated with functions with a large number of parameters, often described as difficult to understand and maintain. In \verb|G2|, there is a recommendation to avoid functions with many parameters, stating that \textit{“if your functions take a very large number of arguments, suspect they should be broken apart, because it's unlikely all arguments are used simultaneously”}.

To illustrate additional cases, we present further examples. In \verb|G1|, the discussion of \textit{Long Method} highlights the decomposition of large functions into smaller, composable units, as it is mentioned that \textit{``imperative style code in one big function ... it's typically better to write a bunch of small functions that each does a specific transformation then compose them declaratively ... Having a bunch of small functions also makes the code easier to reuse''}.

In \verb|G14|, there is evidence of \textit{Shotgun Surgery}, reflected in concerns that \textit{``any change in the representation of those data will have a big impact on the tests code because it will force us to change many tests''}. The same document also points to \textit{Inappropriate Intimacy}, stating that \textit{``this test knows too much about the structure of db''}, indicating excessive coupling to internal details.

Despite these discussions, feedback from the Clojure community indicated that traditional code smells are often considered of limited relevance in this context, with comments such as:

\textit{``The Traditional Smells section lists many code smells that are popular in OOP languages (such as Java). Most of them are not applicable to Clojure... I disagree with nearly every one, and think the entire section could be deleted without losing anything.''} 

\textit{``I think the Clojure section is good and repeats a lot of good advice from sources (Clojure style guide, other linters), the functional section is good albeit generic, and the traditional section can safely be deleted.''} 

\textit{``The whole traditional smells section is of dubious use.''}

\begin{table}[t]
\centering
\caption{Traditional code smells (grey literature)}
\label{tab:traditional-smells}

\begin{tabular}{p{0.42\columnwidth} p{0.38\columnwidth} c}
\toprule
\small
\textbf{Code Smell in Clojure} & \textbf{Documents} & \textbf{\#} \\
\midrule

Long Parameter List & G1, G2 & 2 \\
Long Method & G1 & 1 \\
Shotgun Surgery & G14 & 1 \\
Inappropriate Intimacy & G14 & 1 \\
Comments & G9 & 1 \\

\bottomrule
\end{tabular}

\end{table}

\begin{tcolorbox}[
    colback=gray!5,
    colframe=gray!60,
    boxrule=0.5pt,
    arc=2mm
]
\textbf{RQ1 Answer:} We found evidence that Clojure developers discuss traditional code smells, although not always using the terminology proposed by Fowler. However, community feedback suggests that these smells are generally perceived as less relevant in the Clojure ecosystem.
\end{tcolorbox}

\subsection{RQ2. Do Clojure developers discuss other smells?}
\label{sec:rq2-result}

This work organizes the analyzed code smells into two complementary categories: (i) \textit{Clojure-specific code smells}, directly derived from community discussions and empirical evidence centered on the language’s ecosystem, and (ii) \textit{functional-style code smells}, also identified in the analyzed documents, but which exhibit a more general nature of the functional programming paradigm rather than being specific to the Clojure ecosystem. The complete catalog is publicly available on GitHub~\cite{clj-catalog} and includes descriptions, code examples, evidence sources, and discussion excerpts for each smell\footnote{The catalog is actively maintained and continuously evolving.}. This second category does not constitute a formally established taxonomy nor has it been empirically validated across multiple functional programming languages, and is treated in this work as a set of descriptive evidence of the functional paradigm as observed in the context of Clojure.

\subsubsection{Clojure-specific code smells}

The Table~\ref{tab:clojure_smells_catalog} presents the Clojure-specific code smells considered in this study, along with their descriptions and respective sources of evidence. In general, these findings reflect recurring practices in the use of Clojure and situations in which language constructs are used in a non-idiomatic manner or in disagreement with widely accepted recommendations in the ecosystem, as observed in technical documentation and community discussions.

An example can be observed in the \textit{Unnecessary Macros} smell, which consists of the inappropriate use of macros in contexts where simple functions would be sufficient. In document G8, it appears as follows:

\textit{``Using macros when regular functions would do is a good example of that. It is absolutely possible to write impenetrable Clojure if you start doing weird things just because you can''}.

Regarding the Clojure-specific smells, \textit{Unnecessary Macros} obtained 88.37\% of \textit{Yes} responses (38/43), followed by \textit{Thread Ignorance} and \textit{Nested Forms}, both with 76.74\% (33/43). \textit{Verbose Checks} also obtained 71.43\% of positive responses (30/42). Some participants selected \textit{I don't know} for specific smells, indicating uncertainty or lack of familiarity with particular practices. The complete survey results for all Clojure-specific smells are available in the \textit{Survey Analysis.xlsx} spreadsheet included in the replication package~\cite{replication-package}.

\begin{table*}[t]
\centering
\small
\caption{Clojure-specific code smells, descriptions, and evidence sources}
\label{tab:clojure_smells_catalog}

\begin{tabular}{
>{\raggedright\arraybackslash}p{0.18\textwidth}
>{\raggedright\arraybackslash}p{0.7\textwidth}
>{\centering\arraybackslash}p{0.05\textwidth}
}
\toprule
\textbf{Code Smell} & \textbf{Brief Description} & \textbf{Docs} \\
\midrule

Unnecessary Macros & Using macros instead of simpler language constructs. & G8 \\

Namespaced Keys Neglect & Using unqualified keywords instead of namespaced ones, increasing ambiguity and the risk of key collisions. & G1 \\

Improper Emptiness Check & Using verbose or non-idiomatic checks to test collection emptiness. & G2 \\

Accessing non-existent Map Fields & Accessing map keys without explicit handling of missing values, making it ambiguous whether \texttt{nil} represents absence or an actual value. & G2 \\

Unnecessary \texttt{into} & Using \texttt{into} where more concise or idiomatic alternatives exist, leading to unnecessarily verbose or inefficient code. & G2 \\

Conditional Build-Up & Building state through multiple conditional updates, leading to verbose, imperative-style code. & G2 \\

Verbose Checks & Manually implementing common checks instead of using existing idiomatic predicates, resulting in unnecessarily verbose and less readable code. & G2 \\

Production \texttt{doall} & Using \texttt{doall} in production code to force lazy evaluation, potentially causing memory or performance issues. & G2 \\

Redundant \texttt{do} Block & Wrapping expressions in an explicit \texttt{do} inside constructs that already support implicit sequencing, adding unnecessary verbosity. & G2 \\

Thread Ignorance & Avoiding threading macros in favor of deeply nested forms or repetitive bindings, resulting in verbose and hard-to-follow data flow. & G2 \\

Nested Forms & Unnecessarily nesting binding or iteration forms instead of combining them, increasing structural complexity and reducing readability. & G2 \\

Direct Usage of \texttt{clojure.lang.RT} & Calling internal runtime APIs instead of public Clojure abstractions, leading to fragile and non-portable code. & G5 \\

\bottomrule
\end{tabular}
\end{table*}

\subsubsection{Code smells in functional style}

The second category comprises code smells that emerge as generalizations of the functional programming paradigm. The examples identified in discussions related to Clojure are presented in Table~\ref{tab:functional_smells}. Unlike the previous category, these smells are not treated as Clojure-specific issues, but rather as recurring patterns observed in functional programming practices. Their inclusion in this work should therefore be understood as an interpretative synthesis derived from partial empirical evidence observed in the context of Clojure, rather than as a formally consolidated taxonomy of functional programming smells.

An example appears in discussion G1, where developers mention practices such as “overabundance of partial application and composition”, “designing everything as higher order functions”, and “very deep call stacks” as problematic patterns discussed in the context of Clojure and functional programming.

Among the functional-style smells, \textit{Lazy Side Effects} obtained the highest agreement, with 90.48\% of \textit{Yes} responses (38/42), followed by \textit{Hidden Side Effects} with 78.57\% (33/42) and \textit{Explicit Recursion} with 71.43\% (30/42). In contrast, \textit{Deeply-nested Call Stacks} obtained 48.78\% of positive responses (20/41). The complete survey results for all functional-style smells are available in the \textit{Survey Analysis.xlsx} spreadsheet included in the replication package~\cite{replication-package}.

\begin{tcolorbox}[
    colback=gray!5,
    colframe=gray!60,
    boxrule=0.5pt,
    arc=2mm
]
\textbf{RQ2 Answer}: We found evidence that Clojure developers discuss additional smells beyond traditional code smells, including both Clojure-specific and functional-style smells. Results from the community interaction and practitioner survey indicate that these smells are perceived as more relevant in the ecosystem than traditional smells.
\end{tcolorbox}

\begin{table*}[t]
\centering
\small
\caption{Functional-style code smells, descriptions, and evidence sources}
\label{tab:functional_smells}

\begin{tabular}{
>{\raggedright\arraybackslash}p{0.18\textwidth}
>{\raggedright\arraybackslash}p{0.7\textwidth}
>{\centering\arraybackslash}p{0.05\textwidth}
}
\toprule
\textbf{Code Smell} & \textbf{Brief Description} & \textbf{Docs} \\
\midrule

Trivial Lambda &
Overuse of anonymous functions instead of named functions, reducing clarity and reusability. & G1, G2 \\

Inefficient Filtering &
Overuse of filtering in data generation, producing many invalid values that are discarded instead of generating valid values directly. & G10 \\

Overabstracted Composition &
Excessive function composition or partial application that increases abstraction and reduces readability of data flow. & G1 \\

Deeply-nested Call Stacks &
Excessive nesting of function calls, making control flow harder to follow and debug. & G1 \\

Overuse of Higher-Order Functions &
Excessive use of functions that take or return other functions, increasing abstraction and reducing readability. & G1 \\

Lazy Side Effects &
Side effects executed within lazy evaluation, leading to unpredictable execution behavior. & G12 \\

Hidden Side Effects &
Side effects that are not explicit in function structure or naming, reducing predictability and testability. & G1 \\

Explicit Recursion &
Use of manual recursion instead of higher-level abstractions such as map, reduce, or filter when applicable. & G1 \\

Positional Return Values &
Returning multiple values using positional structures instead of explicit named mappings, reducing clarity and maintainability. & G2 \\

\bottomrule
\end{tabular}
\end{table*}

\subsection{RQ3. Does a well-known static code analysis tool for Clojure detect code smells?}
\label{sec:rq3-result}
In our analysis, only 2 of the 26 cataloged smells are explicitly detected by existing tools: \textit{Improper Emptiness Check}, covered by clj-kondo's \texttt{:not-empty?}, and \textit{Redundant do Block}, which is covered by \texttt{:redundant-do}. A third smell, Verbose Checks, receives partial support via clj-kondo's \texttt{:not-nil?}, capturing a common subcase involving explicit nil comparisons. The remaining smells are not directly supported by the analyzed tools. Overall, the results suggest that current Clojure linters focus primarily on local syntactic and idiomatic issues, while most smells remain unsupported.

\begin{tcolorbox}[
    colback=gray!5,
    colframe=gray!60,
    boxrule=0.5pt,
    arc=2mm
]
\textbf{RQ3 Answer}: Existing static analysis tools for Clojure provide only limited support for the cataloged smells. Among the 26 identified smells, only 2 are detected by current tools.
\end{tcolorbox}

\section{THREATS TO VALIDITY}
\label{sec:threats}

As a replication study, this work inherently shares structural threats with the baseline evaluation conducted for Elixir \cite{Vegi_2022}. The vulnerabilities are as follows:

 \textbf{Construct Validity.} The primary risk involves missing relevant discussions due to search query formulation. To mitigate this, we adopted a pre-calibrated query structure from the baseline study, modifying only the programming language keyword.
 
 \textbf{Conclusion Validity.} Our reliance on technical blogs and community channels introduces non-peer-reviewed data, which risks being influenced by personal preferences rather than genuine design flaws. To counteract this subjective bias during the qualitative classification, all candidate smells were extracted independently by the first two authors and systematically reconciled via joint consensus to ensure only relevant quality concerns were retained.
 
 \textbf{Internal and External Validity.} Generalizing findings from a single grey literature source limits the exhaustiveness of our catalog. We mitigated this by cross-referencing our data with active community hubs to anchor the catalog in real practitioner perceptions. A broader triangulation via repository mining is ongoing.

\section{CONCLUSION AND FUTURE WORK}
\label{sec:future-work}

In this paper, we presented the early findings of a replication study investigating code smells within the Clojure ecosystem, adapting a methodology previously applied to the Elixir language. By reviewing 15 selected documents from the grey literature and conducting an initial community consultation, we gathered preliminary evidence regarding how software quality issues manifest in this modern functional language.

Regarding \textbf{RQ1}, our findings reveal that while traditional code smells are occasionally discussed by developers, the broader Clojure community generally perceives them as less relevant or non-applicable due to the functional nature of the language. For \textbf{RQ2}, our grey literature review uncovered a preliminary set of 12 Clojure-specific smells alongside 9 broader functional-style code smells, which represent recurring non-idiomatic practices in this paradigm. Furthermore, by interacting with practitioners and analyzing their survey responses, we were able to evaluate the perceived relevance of these identifies smells according to the community's perspective. For \textbf{RQ3}, our analysis of existing tools revealed a significant gap in automated code quality assessment within this ecosystem, as current Clojure linters focus predominantly on local syntactic or style issues, leaving the vast majority of our cataloged structural and design-related smells unsupported.

As future work, we plan to expand this investigation in three main directions. First, we intend to continue the replication process by mining open-source Clojure repositories on GitHub, analyzing artifacts such as source code, commits, issues, and pull requests to identify new candidate smells and gather further empirical evidence of their occurrence in production systems. Second, we are currently developing custom static analysis rules and tools tailored to the Clojure smells catalog~\cite{clj-catalog}, with the goal of providing automated support for detecting language-specific smells that still lack diagnostic coverage. As part of this effort, we are also contributing to \textit{clj-kondo} by proposing rules related to the identified smells. Finally, we intend to investigate best practices for removing the identified smells and propose a catalog of refactoring strategies tailored to both Clojure-specific and functional-style code smells.

%...

\section{REPLICATION PACKAGE}
\label{sec:replication-package}

The replication package of this study includes: (i) the complete catalog of identified code smells (\textit{Code Smells Catalog.xlsx}); (ii) the survey questionnaire (\textit{Survey.pdf}); (iii) anonymized survey responses (\textit{Survey Responses.xlsx}); (iv) the quantitative survey analysis (\textit{Survey Analysis.xlsx}); and (v) the list of analyzed Google results (\textit{Google Top-60.xlsx}). The package is publicly available online~\cite{replication-package}.

%...

\section*{Acknowledgements}

We would like to express our gratitude to the members of the Clojure community and the team at Nubank who generously participated in this study, interacted with our repository, and shared their valuable perspectives to validate and enrich our code smells catalog.

\bibliographystyle{ACM-Reference-Format}
\bibliography{sample-base}

@book{fowler1999,
  author = {Fowler, Martin and Beck, Kent and Brant, John and Opdyke, William and Roberts, Don},
  title = {Refactoring: Improving the Design of Existing Code},
  publisher = {Addison-Wesley},
  year = {1999},
  address = {Boston, MA, USA},
  edition = {1st},
  isbn = {0-201-48567-2},
}

@article{Carvalho2019Empirical,
  author = {Carvalho, Suelen Goularte and Aniche, Maurício and Veríssimo, Júlio and Durelli, Rafael S. and Gerosa, Marco Aurélio},
  title = {An empirical catalog of code smells for the presentation layer of Android apps},
  journal = {Empirical Software Engineering},
  volume = {24},
  pages = {3546-3586},
  year = {2019},
  doi = {10.1007/s10664-019-09768-9}
}

@inproceedings{Zhang2022Code,
  author = {Zhang, Haiyin and Cruz, Luís and Deursen, Arie Van},
  title = {Code Smells for Machine Learning Applications},
  booktitle = {Proceedings - 1st International Conference on AI Engineering - Software Engineering for AI, CAIN 2022},
  year = {2022},
  pages = {217--228},
  address = {Pittsburgh, PA, USA},
  publisher = {IEEE/ACM},
  doi = {10.1145/3522664.3528620}
}

@inproceedings{Vegi_2022, series={ICPC ’22},
   title={Code smells in Elixir: early results from a grey literature review},
   url={http://dx.doi.org/10.1145/3524610.3527881},
   DOI={10.1145/3524610.3527881},
   booktitle={Proceedings of the 30th IEEE/ACM International Conference on Program Comprehension},
   publisher={ACM},
   author={Vegi, Lucas Francisco da Matta and Valente, Marco Tulio},
   year={2022},
   month=May, pages={580–584},
   collection={ICPC ’22} 
}

@article{francisco2023understanding,
  author = {Vegi, Lucas Francisco Matta and Valente, Marco Tulio},
  title = {Understanding code smells in {E}lixir functional language},
  journal = {Empirical Software Engineering},
  volume = {28}, 
  number = {102},
  pages = {1-32},
  doi = {10.1007/s10664-023-10343-6},
  year = {2023}
}

@techreport{haskellSmellsCowie2005,
  title={Detecting bad smells in {H}askell},
  author={Cowie, Jonathan},
  institution = {University of Kent, UK},
  year={2005}
}

@INPROCEEDINGS{ErlangModularitySmells,
  author={Li, Huiqing and Thompson, Simon},
  booktitle={10th IEEE Working Conference on Source Code Analysis and Manipulation (SCAM)}, 
  title={Refactoring Support for Modularity Maintenance in {E}rlang}, 
  year={2010},
  pages={157-166},
  doi={10.1109/SCAM.2010.17}
}

@inproceedings{CodeCloneErlang01,
    author = {Li, Huiqing and Thompson, Simon},
    title = {Clone Detection and Removal for {E}rlang/{OTP} within a Refactoring Environment},
    year = {2009},
    doi = {10.1145/1480945.1480971},
    booktitle = {ACM SIGPLAN Workshop on Partial Evaluation and Program Manipulation (PEPM)},
    pages = {169-178}
}

@inproceedings{CodeCloneErlang02,
    author = {Seijas, Pablo Lamela and Thompson, Simon},
    title = {Identifying and Introducing Interfaces and Callbacks Using {W}rangler},
    year = {2016},
    doi = {10.1145/3064899.3064909},
    booktitle = {28th Symposium on the Implementation and Application of Functional Programming Languages (IFL)},
    pages = {1-13}
}

@article{CodeCloneErlang03,
  title={The pragmatics of clone detection and elimination},
  author={Thompson, Simon and Li, Huiqing and Schumacher, Andreas},
  journal={The Art, Science, and Engineering of Programming},
  year={2017},
  volume={1},
  number={2},
  pages={1-34}
}

@article{CodeCloneErlang04,
    author = {F\"{o}rd\H{o}s, Vikt\'{o}ria and T\'{o}th, Melinda},
    title = {Identifying Code Clones with {R}efactor{E}rl},
    year = {2016},
    volume = {22},
    number = {3},
    doi = {10.14232/actacyb.22.3.2016.1},
    journal = {Acta Cybernetica},
    pages = {553–571}
}

@inproceedings{CodeCloneHaskell01,
    author = {Brown, Christopher M. and Thompson, Simon},
    title = {Clone Detection and Elimination for {H}askell},
    year = {2010},
    doi = {10.1145/1706356.1706378},
    booktitle = {ACM SIGPLAN Workshop on Partial Evaluation and Program Manipulation (PEPM)},
    pages = {111–120}
}

@inproceedings{hickey2008clojure,
author = {Hickey, Rich},
title = {The Clojure programming language},
year = {2008},
isbn = {9781605582702},
publisher = {Association for Computing Machinery},
address = {New York, NY, USA},
url = {https://doi.org/10.1145/1408681.1408682},
doi = {10.1145/1408681.1408682},
booktitle = {Proceedings of the 2008 Symposium on Dynamic Languages},
articleno = {1},
numpages = {1},
location = {Paphos, Cyprus},
series = {DLS '08}
}

@article{lisp,
  title={Lisp},
  author={Winston, Patrick Henry and Horn, Berthold K},
  year={1986},
  publisher={Addison Wesley Pub., Reading, MA}
}

@article{garousi2019,
  author    = {Vahid Garousi and Michael Felderer and Mika V. Mäntylä},
  title     = {Guidelines for including grey literature and conducting multivocal literature reviews in software engineering},
  journal   = {Information and Software Technology},
  volume    = {106},
  pages     = {101--121},
  year      = {2019},
  doi       = {10.1016/j.infsof.2018.09.006},
  publisher = {Elsevier}
}

@misc{clojure_official,
  title        = {Clojure Programming Language},
  author       = {Rich Hickey},
  year         = {2026},
  howpublished = {\url{https://clojure.org/}},
  note         = {Accessed: 2026-05-14}
}

@misc{clj_kondo,
  author = {{clj-kondo Contributors}},
  title = {clj-kondo},
  year         = {2026},
  howpublished = {\url{https://github.com/clj-kondo/clj-kondo}},
  note = {Accessed: 2026-05-14}
}

@misc{eastwood,
  author = {{Eastwood Contributors}},
  title = {Eastwood: Clojure Linter},
  year         = {2024},
  howpublished = {\url{https://github.com/jonase/eastwood}},
  note = {Accessed: 2026-05-14}
}

@misc{joker,
  author       = {{joker Contributors}},
  title        = {Joker: Small Clojure Interpreter, Linter, and Formatter},
  year         = {2026},
  howpublished = {\url{https://github.com/candid82/joker}},
  note         = {Accessed: 2026-05-19}
}

@misc{kibit,
  author = {{Kibit Contributors}},
  title = {Kibit},
  year         = {2024},
  howpublished = {\url{https://github.com/clj-commons/kibit}},
  note = {Accessed: 2026-05-14}
}

@misc{clojure_lsp,
  author = {{clojure-lsp contributors}},
  title = {Clojure LSP},
  year         = {2024},
  howpublished = {\url{https://clojure-lsp.io/}},
  note = {Accessed: 2026-05-14}
}

@misc{sonarqube,
  author       = {{SonarSource}},
  title        = {{SonarQube: Continuous Inspection of Code Quality}},
  year         = {2024},
  howpublished = {\url{https://www.sonarqube.org}},
  note         = {Accessed: 2026-05-14}
}

@misc{pmd,
  author       = {{PMD Developers}},
  title        = {{PMD: Source Code Analyzer for Java, JavaScript, Salesforce, and More}},
  year         = {2024},
  howpublished = {\url{https://pmd.github.io}},
  note         = {Accessed: 2026-05-14}
}

@inproceedings{ctd,
 author = {Lucas Vegi and Marco Valente},
 title = { Code Smells and Refactorings for Elixir},
 booktitle = {Anais do XXXVIII Concurso de Teses e Dissertações},
 location = {Maceió/AL},
 year = {2025},
 issn = {2763-8820},
 pages = {94--103},
 publisher = {SBC},
 address = {Porto Alegre, RS, Brasil},
 doi = {10.5753/ctd.2025.9280},
 url = {https://sol.sbc.org.br/index.php/ctd/article/view/36357}
}

@book{brooks1982,
  author = {Frederick P. Brooks Jr.},
  title = {The Mythical Man-Month: Essays on Software Engineering},
  publisher = {Addison-Wesley},
  address = {Reading, Massachusetts},
  year = {1982}
}

@ARTICLE{1265817,
  author={Mens, T. and Tourwe, T.},
  journal={IEEE Transactions on Software Engineering}, 
  title={A survey of software refactoring}, 
  year={2004},
  volume={30},
  number={2},
  pages={126-139},
  doi={10.1109/TSE.2004.1265817}}

@inproceedings{harris2005composable,
author = {Harris, Tim and Marlow, Simon and Peyton-Jones, Simon and Herlihy, Maurice},
title = {Composable memory transactions},
year = {2005},
isbn = {1595930809},
publisher = {Association for Computing Machinery},
address = {New York, NY, USA},
url = {https://doi.org/10.1145/1065944.1065952},
doi = {10.1145/1065944.1065952},
booktitle = {Proceedings of the Tenth ACM SIGPLAN Symposium on Principles and Practice of Parallel Programming},
pages = {48–60},
numpages = {13},
location = {Chicago, IL, USA},
series = {PPoPP '05}
}

@misc{mosley2006out,
  title={Out of the Tar Pit},
  author={Moseley, Ben and Marks, Peter},
  year={2006},
  month={feb},
  note={Available at: \url{https://curtclifton.net/papers/MoseleyMarks06a.pdf}}
}

@techreport{bagwell2001ideal,
  title={Ideal Hash Trees},
  author={Bagwell, Phil},
  institution={EPFL},
  year={2001},
  note={Available at: \url{https://infoscience.epfl.ch/server/api/core/bitstreams/f66a3023-2cd0-4b26-af6e-91a9a6ae7450/content}}
}

@misc{clojure_reader,
  author = {{Clojure.org}},
  title = {The Reader},
  year = {2026},
  url = {https://clojure.org/reference/reader},
  note = {Accessed: 2026-05-18}
}

@dataset{replication-package,
  author    = {Walber Araújo and José Truta and Lucas Vegi and Marco Tulio Valente and João Brunet},
  title     = {Replication Package: Code Smells in the Clojure Ecosystem},
  year      = {2025},
  publisher = {Zenodo},
  doi       = {10.5281/zenodo.20293948},
  url       = {https://doi.org/10.5281/zenodo.20293948}
}

@misc{clj-catalog,
author = {Walber Wesley Félix de Araújo Filho and José do Bomfim Truta Neto and Rafael Antônio de Lucena Serey and Thiago Alves Laurentino and João Arthur Brunet Monteiro},
title = {clj-smells-catalog: Catalog of Clojure-related code smells},
year = {2025},
howpublished = {\url{https://github.com/nufuturo-ufcg/clj-smells-catalog}},
note = {NuFuturo - UFCG. Accessed: May 20, 2026}
}

\end{document}